\documentclass[12pt, a4paper]{amsart}
\usepackage{amsmath,amssymb,amsfonts}
\usepackage{graphicx}
\usepackage[scaled=1.0]{newtxtext,newtxmath}
\usepackage{hyperref}
\usepackage{bbm}
\usepackage{multirow}
\usepackage{bm}

\usepackage{tikz}
\usetikzlibrary{arrows}
\usetikzlibrary{arrows.meta}
\tikzset{
	tipA/.tip={Triangle[angle=45:10pt]}	
}

\usepackage{makecell}

\usepackage[english]{babel}

\usepackage[colorinlistoftodos]{todonotes}

\usepackage{scalerel}
\newlength\bshft
\def\mybold#1{\ThisStyle{\ooalign{$\SavedStyle#1$\cr%
			\kern-\bshft$\SavedStyle#1$\cr%
			\kern\bshft$\SavedStyle#1$}}}

\newcommand\pbmbb[1]{ {\mybold{\bm{\mathbb{#1}}}} } 
\newcommand\phase[1]{ \vvmathbb{#1} }
\newcommand\bphase[1]{\overline{\vvmathbb{#1}}}
\newcommand{\olasep}{odd mLPASEP}
\newcommand{\elasep}{even mLPASEP}
\newcommand{\ds}{\displaystyle}

\title[The phase diagram for a multispecies left-permeable ASEP]
{The phase diagram for a multispecies left-permeable asymmetric exclusion process}

\author{Arvind Ayyer}
\address{Arvind Ayyer, Department of Mathematics, Indian Institute of Science, Bangalore - 560012, India.}
\email{arvind@iisc.ac.in}
\author{Caley Finn} 
\address{Caley Finn, 
	Australian Research Council Centre of Excellence for Mathematical and Statistical Frontiers (ACEMS)\\
	School of Mathematics and Statistics,
	The University of Melbourne \\
	VIC 3010,
	Australia}
\email{cfinn@unimelb.edu.au}
\author{Dipankar Roy}
\address{Dipankar Roy, Department of Mathematics, Indian Institute of Science, Bangalore - 560012, India.}
\email{dipankarroy@iisc.ac.in}
\date{\today}

\begin{document}

\begin{abstract}
We study a multispecies generalization of a left-permeable asymmetric exclusion process (LPASEP) in one dimension with open boundaries. We determine all phases in the phase diagram using an exact projection to the LPASEP solved by us in a previous work. 
In most phases, we observe the phenomenon of dynamical expulsion of 
one or more species. 
We explain the density profiles in each phase using interacting shocks. This explanation is corroborated by simulations.
\end{abstract}


\maketitle

\section{Introduction} 
\label{sec:intro}
The asymmetric simple exclusion process (ASEP) serves as a paradigmatic model for understanding the phenomena of nonequilibrium transport. In the open ASEP in one-dimension, particles hop along a finite one-dimensional lattice connected to reservoirs with asymmetric hopping rules and excluded volume interactions. 
The nonequilibrium steady state (NESS) of the totally asymmetric variant (TASEP) was computed exactly using a new technique called the matrix ansatz by Derrida, Evans, Hakim and Pasquier~\cite{derrida1993}. They derived the phase diagram for the TASEP through the exact computation of macroscopic quantities such as current and density in the NESS. 
The matrix ansatz has since then been successfully used to obtain the NESS of other one-dimensional lattice gases; see ~\cite{blythe2007} for a review.

While the ASEP with a single species of particle is relatively well-understood, the problem of open ASEPs with multiple species is considerably open. This is not a purely academic exercise.
Open ASEPs with one or more than one species have found applications in studies pertaining to traffic flow \cite{schadschneider2002}, biological systems \cite{chowdhury2000} and cell motility \cite{penington2011} despite its simple dynamical rules. 
Some exactly solvable open ASEPs with two species are known~\cite{Evans1995, Arita2006b, uchiyama2008, ayyer2009, ayyer2012, crampe2015, crampe-evans-mallick-ragoucy-vanicat2016, ayyer2018}.
On the other hand, there has been a lot of progress in understanding the NESS of closed ASEPs (i.e. ASEPs with periodic boundary conditions) with arbitrary number of species. The matrix ansatz has been successfully developed for both the multispecies TASEP~\cite{evans-ferrari-mallick-2009} and for the multispecies ASEP~\cite{prolhac-evans-mallick-2009}. The NESS for the multispecies TASEP has also been obtained using queueing-theoretic techniques in \cite{FM06, FM07}. Using the latter, nearest-neighbour $3$-point correlations as well as arbitrary $2$-point correlations have been computed in \cite{ayyer-linusson-2016}.

For the open ASEP with arbitrary number of species, the only exactly solved model so far is the so-called {\em mASEP}. It was first considered by Cantini, Garbali, de Gier and Wheeler, who proved a formula for the nonequilibrium partition function in \cite{cantini2016}. The phase diagram of the mASEP was understood by the first and the third author in~\cite{ayyer2017} using exact projections by the so-called {\em colouring} technique. This idea was earlier used to understand the phase diagram of open two-species ASEPs in \cite{ayyer2009, ayyer2012, crampe-evans-mallick-ragoucy-vanicat2016}.
We should also mention that integrable boundary rates for multispecies open ASEPs have been classified in \cite{crampe2016}.

In the present work, we derive the exact phase diagram for a model with arbitrary number of species generalizing the left-permeable ASEP (LPASEP) introduced by us in an earlier work~\cite{ayyer2018}. By analogy with the mASEP, we call the model the mLPASEP. We use exact colourings of the mLPASEP to the LPASEP to determine the multidimensional phase diagram. We also give physical explanations for the currents and densities in each phase by appealing to the shock picture. We observe the phenomenon of {\em dynamical expulsion} here too just as in the mASEP~\cite{ayyer2017}.
There are two kinds of colourings depending on the parity of the number of species. We focus on the technically easier case of the \olasep{} for the most part. The differences for the \elasep{} are highlighted in Appendix~\ref{sec:ELASEP}.

We note that we do not have a matrix ansatz for the mLPASEP. Instead, the colouring method allows us to project to the LPASEP where there is a matrix ansatz. This allows us to derive the phase diagram exactly. 
This colouring technique works both for the finite system and the one in the thermodynamic limit, but it is most efficient for calculating simple correlations like the density and current. Calculating higher order correlations this way will be quite difficult. Determining the complete steady state is possible in principle, but intractable in practice. It would be interesting to determine the steady state of the mLPASEP using a matrix ansatz.

The plan of the article is as follows. We define the models in Section~\ref{sec:def}. Since the understanding of the phase diagram of the \olasep{} depends crucially on the LPASEP, we review the latter in Section~\ref{sec:irasep}. 
In Section~\ref{sec:genpd}, we discuss the colouring approach and derive the exact phase diagram of the \olasep{} in the thermodynamic limit. We also give formulas for the currents and densities in all phases. 
To illustrate the ideas, we explain the three dimensional phase diagram of the \olasep{} with $5$ species in Section~\ref{sec:pd-r=2}. 
We also perform simulations of the NESS for this case in Figure~\ref{fig:Asimd}.
Lastly, we explain the coarse features of the density profiles in Section~\ref{sec:gshockpicture} by appealing to the generalized shock picture. Here, we simulate the shocks on certain phase boundaries in Figure~\ref{fig:5asepcg} and study the spatio-temporal evolution of the shock in Figure~\ref{fig:5asepshocksim} for the \olasep{} with $5$ species.

\section{Model definitions}
\label{sec:def}

The {\em asymmetric exclusion process} or ASEP is an interacting particle system or lattice gas defined on a (finite or infinite) lattice, where each site is occupied by at most one particle. The dynamics of the ASEP is stochastic and in continuous-time. 
For our purposes, the ASEP will be defined on a finite one-dimensional lattice of size $L$.

The {\em multispecies left-permeable ASEP} or mLPASEP is a variant of the ASEP where there are several different types or {\em species} of particles. Each site of the lattice is occupied by exactly one particle of a certain species. In our convention, the vacancies too are considered as a species of particles. The model with odd (resp. even) number of species is referred to as the {\em \olasep{}} (resp. {\em \elasep{}}).  
We now give the precise definitions of the models.

\subsection{The \olasep{}: mLPASEP with $(2r+1)$-species}\label{ssec:olasep}
Each species in the \olasep{} is labelled by an element of $\pbmbb{L} := \left\lbrace \overline{r},\dots,  \overline{1},0,1, \dots \allowbreak ,r \right\rbrace $. The barred labels should be regarded as negative integers with the natural order relation : $\overline{r} < \cdots < 0 < \cdots < r$. The dynamics is as follows.
In the bulk, the rules for exchange of particles $i$ and $j$ $(i,j \in \pbmbb{L})$ between two neighbouring sites are given by
\begin{equation}
	ij \rightarrow ji  \quad \text{with rate} 
	\begin{cases} 
		1 \quad \text{if} \ i>j,  \\
		q \quad \text{if} \ i<j,
	\end{cases} 
	\label{eq:bulk}
\end{equation}
where we impose $q<1$. 
At the left boundary, either of the two type of transitions are permissible: (i) a species can replace a smaller species, or (ii) a species whose label is nonnegative can replace a higher order species.
 These transitions and corresponding rates are summarized as 
\begin{equation}
	i \rightarrow  j \quad \text{with rate} 
	\begin{cases} 
		\alpha_{j} \quad \text{if} \ \overline{r} \leq i < j,  \\
		\gamma_{j}\quad \text{if} \ i>j\geq 0.
	\end{cases}
	\label{eq:olb}
\end{equation}
The rates $\alpha_{j}$ are independent positive parameters, whereas the $\gamma_j$'s are defined in terms of the $\alpha_{j}$'s and $q$. 
To write the relation concisely, we define the quantities $\theta_{j} = \sum_{i=j}^{r} \alpha_{i}$, and $\phi_{j} = \sum_{i= \overline{j-1}}^{j-1} \alpha_{i}$ for $j \in [r]$. Then
\begin{equation}
\gamma_{j} = 
\begin{cases}
\ds	\frac{ \phi_{1} \left( \phi_{1} +\theta_{1} -1 +q  \right) }{ \phi_{1} +\theta_{1} }  
		& \quad \text{if}  \ j = 0 ,  \\[0.4cm]
\ds \alpha_{j} + \alpha_{\overline{j}} - \frac{(1-q) \left( \alpha_{j} \phi_{j} + \alpha_{j}\theta_{j} + \alpha_{\overline{j}} \theta_{j} \right) }{(\theta_{j} + \phi_{j})( \theta_{j+1} + \phi_{j+1})}
		
		& \quad \text{if}  \ j > 0 .
	\end{cases}  \label{eq:gammadef}
\end{equation}
This precise functional form of $\gamma_{j}$'s is necessary to be able to appeal to the integrability of the LPASEP in constructing the phase diagram of the mLPASEP. A similar choice of function was necessary there in order to construct the exact steady state using the matrix ansatz. See~\cite{ayyer2018} for more details.
In order that $\gamma_{j}$'s are positive, we impose the restriction $\theta_0 > 1 -q $. 
At the right boundary, an unbarred species $i$ can replace or be replaced by its barred counterpart with the following rates
\begin{equation}
	\begin{cases}
	i \rightarrow  \overline{i} \quad \text{with rate} \ \beta , \\
	\overline{i}  \rightarrow i  \quad \text{with rate} \ \delta , 
	\end{cases} \label{eq:rb}
\end{equation}
where $\beta$ and $\delta$ are positive parameters. Thus, species $0$ can neither enter nor exit from the right boundary.

\subsection{The \elasep{}: mLPASEP with $(2r)$-species} 
\label{sec:defB}
The label set for all species in \elasep{} is 
$\pbmbb{L}_{0} \equiv \pbmbb{L} \setminus \left\lbrace 0 \right\rbrace $. The bulk and right boundary transitions given by \eqref{eq:bulk} and \eqref{eq:rb} are unaltered. The left boundary transitions resemble those in the \olasep{}, except that species $\overline{1}$ can also replace a higher species. More precisely,
\begin{equation}
i \rightarrow  j \quad \text{with rate} 
\begin{cases} 
	\alpha_{j} \quad \text{if} \ \overline{r} \leq i<j,  \\
	\gamma'_{j}\quad \text{if} \ i>j\geq 1,	\\
	\gamma'_{0} \quad \text{if} \ i>j=\overline{1} .
\end{cases}
\label{eq:elb}
\end{equation}
As before, the rates $\alpha_{j}$ are independent positive parameters and the $\gamma'_j$'s are defined in terms of the $\alpha_{j}$'s and $q$. 
Define $\theta_j$'s as for the \olasep{} and $\phi'_{j} = \sum_{i=1}^{j-1}(\alpha_{i} + \alpha_{\overline{i}})$ for $2\leq k \leq r$. Then 
\begin{equation}
\gamma'_{j} = 
\begin{cases}
\ds	\frac{\phi'_{1}( \phi'_{1} +\theta_{1} - 1+ q)}{ \phi'_{1} +\theta_{1} } -\gamma'_{0} 
		& \quad \text{if}  \ j = 1,  \\[0.4cm]
\ds \alpha_{j} + \alpha_{\overline{j}} - \frac{(1-q) \left( \alpha_{j} \phi'_{j} + \alpha_{j}\theta_{j} + \alpha_{\overline{j}} \theta_{j} \right) }{(\theta_{j} + \phi'_{j})( \theta_{j+1} + \phi'_{j+1})}
		& \quad \text{if}  \ j > 1,
	\end{cases}  \label{eq:egammadef}
\end{equation}
where the rates are chosen so that $\phi'_{1} +\theta_{1}  >  1 - q $ and $\gamma'_{0} < \phi'_{1}( \phi'_{1} +\theta_{1} - 1+ q)/( \phi'_{1} +\theta_{1} )$. 
Again, the reason for this specific functional form of $\gamma'_{j}$'s is to make a connection with the integrability of the LPASEP~\cite{ayyer2018}.

As mentioned above, we will focus on the \olasep{} throughout the article. The treatment of the \elasep{} follows very similar lines and we relegate that discussion to Appendix~\ref{sec:ELASEP}.

\section{The exact solution of the LPASEP}
\label{sec:irasep}
To derive the exact phase diagram of the \olasep{}, we will use the exact solution of the LPASEP \cite{ayyer2018}, which we now recall. 
The LPASEP is the \olasep{} with $r=1$ with slight change of terminology. The particles labelled $\overline{1}$, 0 and 1 in our notation were referred to as {\em vacancies}, {\em first class particles} and {\em second class particles} respectively. 
The transition are given by \eqref{eq:bulk}, \eqref{eq:olb} and \eqref{eq:rb} with $r=1$. For completeness, we record the boundary transitions here:
\begin{center}
	\begin{tabular}{ll}
		\underline{Left}:
		&
		$\begin{cases}
		\overline{1}, 0 \rightarrow 1 \quad \text{ with rate } \alpha_{1},	\\
		\overline{1} \rightarrow 0 \quad \text{ with rate } \alpha_{0},  \\
		1 \rightarrow 0 \quad \text{ with rate } \gamma_{0} ,
		\end{cases}$\\
		&\\
		\underline{Right}:
		&$
		\begin{cases}1 \rightarrow \overline{1} \quad \text{ with rate } \beta, \\
		\overline{1} \rightarrow 1 \quad \text{ with rate } \delta.
		\end{cases}$
	\end{tabular}
\end{center}
The rate $\gamma_{0}$ is dependent on the rates $\alpha_{0}$ and $\alpha_{1}$ via the relation $\gamma_{0} = \alpha_{0} (\alpha_{0} + \alpha_{1} -1 +q) / (\alpha_{0} + \alpha_{1})$. The boundary parameters that determine the phase diagram are $\lambda = \alpha_{0} / \alpha_{1}$ for the left boundary and $b= \kappa^{+}_{\beta,\delta}$ for the right boundary, where
\begin{equation}
\label{def-kappa}
	\kappa^{\pm}_{u,v} = \frac{1-q -u+v \pm \sqrt{(1-q -u+v)^2 + 4u v}}{2 u}. 
\end{equation}

\begin{figure}[htbp!]
		\begin{center}	
			\begin{tikzpicture}[scale=0.5]
			\draw[-{Latex[length=3.5mm]},thick] (0,0)--(0,10);
			\draw[-{Latex[length=3.5mm]},thick] (0,0)--(10,0);
			\draw[black,thick] (5,5)--(10,10);
			\draw[black,thick] (0,0) rectangle (5,5);
			
			\node at (2.5,2.5) {$\bphase{1}$};
			\node at (7.5,2.5) {$\phase{0}$};
			\node at (2.5,7.5) {$\phase{1}$};
			
			\node at (-0.65,0) {$0$};
			\node at (-0.65,5) {$1$};
			\node at (0,10.65) {$b$};
			
			\node at (0,-0.65) {$0$};
			\node at (5,-0.65) {$1$};
			\node at (10.5,0) {$\lambda$};
			
			\node at (10.25,10.625) {$b=\lambda$};

			\end{tikzpicture}
		\end{center}
	\caption{The phase diagram for the LPASEP. See Table~\ref{table:cdlpasep} for details of currents and densities in each phase.}
	\label{fig:lasepphdiag}
\end{figure}
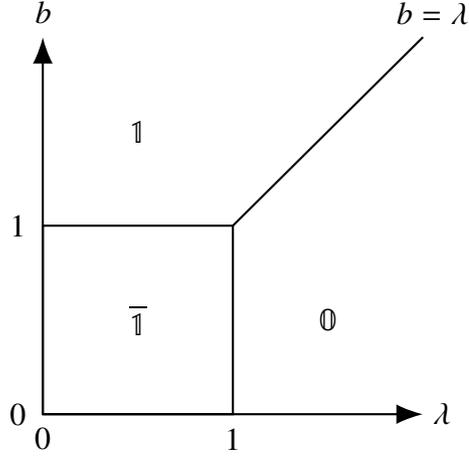

The phase diagram for the steady state of the LPASEP depends on $\lambda$ and $b$ and is plotted in Figure~\ref{fig:lasepphdiag}. It has three phases: phase $\phase{1}$ or high density (HD) phase, phase $\phase{0}$ or low density (LD) phase, and phase $\bphase{1}$ or maximal current (MC) phase. In Table~\ref{table:cdlpasep}, we list the currents and bulk densities in all the three phases and the $\phase{0}-\phase{1}$ co-existence line.

\begin{table}[htbp!]
\renewcommand*{\arraystretch}{1.6}
	\begin{tabular}{|c|c|c|c|c|}
		\hline	
	 Phase & Phase Region &  $\rho_{1}$&  $\rho_{0}$ & $J_{1}$ \\
		\hline
		$\phase{1}$ (HD)	 &	$\max \left\lbrace  \lambda ,1 \right\rbrace < b $ & $\frac{b}{1+b}$ &  $ 0$ & $ (1-q) \frac{b}{(1+b)^2} $\\ \hline
		$\phase{0}$ (LD)	 &  $\max \left\lbrace b,1 \right\rbrace < \lambda $ & $\frac{1}{1+\lambda}$ & $\frac{\lambda -1}{ 1+ \lambda } $ & $ (1-q) \frac{\lambda}{ (1+ \lambda)^2} $ \\ \hline
		$\bphase{1}$ (MC)   &  $\max \left\lbrace \lambda ,b \right\rbrace < 1 $  &  $\frac{1}{2}$ & $ 0$ &$\frac{1}{4}(1-q)$ \\ 
		\hline 
		$\phase{0} - \phase{1}$ co-existence line & $ 1< b = \lambda $ & $\frac{1-x+x b}{1+b}$ & $\frac{(1-x)(b-1) }{1+b}$ &  $ (1-q) \frac{b}{ (1+ b)^2} $\\
		\hline
	\end{tabular} 
	\vspace{0.4cm}
\caption{Currents and bulk densities for the LPASEP. The density and current of $\overline{1}$'s can be determined from the equations 
$\rho_{1} + \rho_{0} +\rho_{\overline{1}} =1$ and $J_{\overline{1}} = - J_1$. 
The normalized site position $x$ equals $i/L$ for site $i$ in the system of size $L$.}
\label{table:cdlpasep}
\end{table}

The coarse features of the density profiles in the steady state in all phases can be explained by a shock picture, just as for the semipermeable ASEP~\cite{ayyer2009}, which we now explain. In the absence of correlations (which we expect in the thermodynamic limit), the current of $1$'s is given by $J_1 = (1-q)\rho_1 (1-\rho_1)$, whereas that of $\overline{1}$'s is $J_{\overline{1}} = (1-q) \rho_{\overline{1}} (1-\rho_{\overline{1}})$ in the opposite direction. Equating these two, we find two solutions:  
\begin{equation}
\label{dens-pairs}
\text{either } \rho_1 = \rho_{\overline{1}} \text{ and } \rho_0 = 1-2\rho_1, \quad \text{ or } \rho_1 = 1-\rho_{\overline{1}} \text{ and } \rho_0 = 0.
\end{equation}
We therefore expect to find this property at all normalized site positions $x = i/L$ in the thermodynamic limit.

\begin{figure}[htbp!]
	\begin{center}	
		\begin{tikzpicture}[scale=0.5]
		\draw[black,thick] (0,0) rectangle (10,10);
		\draw[black] (0,3)--(6,3)--(6,7);
		\draw[black] (0,7)--(10,7);
		\node[] at (-1.5,3){$\scriptstyle 1/(1+b)$};
		
		\node[] at (11.5,7){$\scriptstyle b/(1+b)$}; 
		
		\node[] at (5,8.5){$\overline{1}$};
		\node[] at (7.5,5){$1$};\node[] at (3,5){$0$};
		\draw[<->] (5.5,5)--(6.5,5);
		
		\node at (-1.8,5) {$\rho$};
		\node[] at (0,-0.5){$0$};\node[] at (10,-0.5){$1$};
		\node[] at (-0.5,0){$0$}; \node[] at (10.55,10){1};
		\node[] at (5,-1){$x$};
		\end{tikzpicture}
	\end{center}
\caption{The shock picture for the LPASEP on the co-existence line $b=\lambda >1$. At each normalized position $x = i/L$, the height of a region equals bulk density of the species which the region is labelled with. }
\label{fig:lpasepshock}
\end{figure}
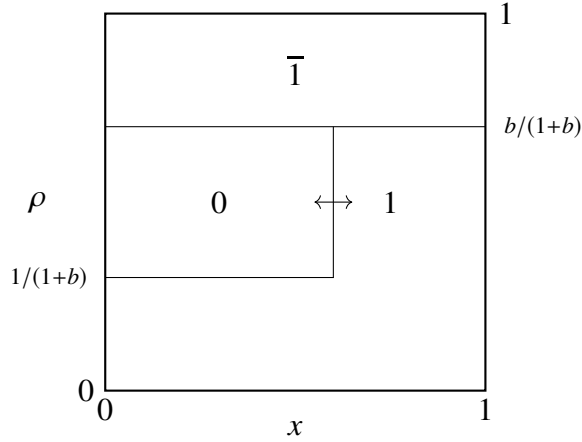

In the LPASEP, a shock is formed between particles of species 0 and species 1, and particles of species $\overline{1}$ act as spectators. This is easiest to explain on the $\phase{1} - \phase{0}$ boundary and is shown in Figure~\ref{fig:lpasepshock}. 
Particles of species 0 and species 1 have discontinuous densities across the shock line. The shock has zero drift and performs a symmetric random walk within the system, leading to linear density profiles for $0$ and $1$. 

In phase~$\phase 1$ (the {\em high density} or HD phase), the shock has negative drift and gets pinned to the left of the system. Therefore, the $0$'s have zero density in the system leading to a high density of $1$'s in the system. This phenomenon is known as {\em dynamical expulsion}~\cite{ayyer2017}, where the boundary parameters cause a species to be absent in the bulk of the system only in certain phases.
In phase~$\phase 0$ (the {\em low density} or LD phase), the shock has positive drift and gets pinned to the right of the system. In that case, the densities of $1$'s and $\overline{1}$'s become equal and constant and the density of $0$'s is non-zero.
In phase~$\bphase 1$ (the {\em maximum current} or MC phase), the height of the shock becomes zero and $0$'s are again dynamically expelled from the system. Here, the density of $1$'s and $\overline{1}$'s become $1/2$, leading to the maximum possible current.

\section{Exact Phase diagram for \olasep{}}
\label{sec:genpd}
To derive the phase diagram, we will construct a series of exact projections from the \olasep{} to the LPASEP. This strategy was successfully used to derive the exact phase diagram for the mASEP \cite{ayyer2017}.
The projection is called the {\em $k$-colouring} and is explained below. We will use it to calculate densities and currents for all species of particles in all phases. 
The reader interested in seeing a concrete example should go to Section~\ref{sec:pd-r=2}, where these results are illustrated for $r=2$. 

The idea of colouring is that some species of particles will be indistinguishable and the dynamics will be the same as that of an \olasep{} with fewer number of species. We emphasize that the colouring is exact in the sense that the projection respects the dynamics both in the bulk and on the boundaries.

To be precise, we fix $k$ between $1$ and $r$. We then identify species $\overline{r}, \ldots , \overline{k},$ as a new species which we label $\overline{1}_{k}$. Similarly, species $k, \ldots , r,$ are identified as $1_{k}$, and  species $\overline{k-1}, \ldots , k-1$, are called $0_{k}$. Each $k$-colouring maps \olasep{} onto the LPASEP with the boundary rates given by:
\begin{center}
	\begin{tabular}{ll}
		\underline{Left}:
		&$
		\begin{cases}
		\overline{1}_{k}, 0_{k} \rightarrow 1_{k} \quad \text{ with rate } \theta_{k},	\\
		\overline{1}_{k} \rightarrow 0_{k} \quad \text{ with rate } \phi_{k}, \\
		1_{k} \rightarrow 0_{k} \quad \text{ with rate } \zeta_{k}, 
		\end{cases}$\\
		&\\
		\underline{Right}:
		&$\begin{cases} 1_{k} \rightarrow \overline{1}_{k} \quad \text{ with rate } \beta, \\
		\overline{1}_{k} \rightarrow 1_{k} \quad \text{ with rate } \delta,
		\end{cases}$
	\end{tabular}
\end{center}
where $\zeta_{k} = \sum_{i=0}^{k-1} \gamma_{i}$. It can be easily checked using the definition of $\gamma_{i}$'s that $ \zeta_{k}= \phi_{k} (\theta_{k} +\phi_{k} -1 +q)/ (\theta_{k} + \phi_{k})$. The left boundary parameter $\lambda_{k} = \theta_{k} /\phi_{k}$ and the right boundary parameter $b = \kappa_{\beta, \delta}^{+}$ (which is independent of $k$) determine part of the phase diagram of the \olasep{}. Since there are such $r$ possible colourings, the overall phase diagram of the generalized model depends on the following $r+1$ parameters: $\lambda_{1},  \dots,  \lambda_{r}$  and $b$.  Note that $\lambda_{1}< \cdots < \lambda_{r}$ by definition.

\begin{figure}[htbp!]
	\begin{center}
		\begin{tikzpicture}[scale=0.35]
		\draw[-{Latex[length=3.5mm]},thick] (0,0)--(0,20);
		\draw[-{Latex[length=3.5mm]},thick] (0,0)--(20,0);
		\draw[black,thick] (0,0) rectangle (3,8);
		\draw[black,thick] (0,0) rectangle (6,8);
		\draw[black,thick] (0,0) rectangle (9,8);
		\draw[black,thick] (0,0) rectangle (12,8);
		\draw[black,thick] (0,0) rectangle (15,8);
		
		\draw[black,thick] (3,8)--(7.5,20); \node at (8,20.75) {$b=\lambda_r$}; \node at (12,20.75) {$\cdots$};
		\draw[black,thick] (6,8)--(15,20);  \node at (16,20.75) {$b=\lambda_{j+1}$};
		\draw[black,thick] (9,8)--(20,17.778); \node at (22.5,17.778) {$b=\lambda_{j}$};
		\draw[black,thick] (12,8)--(20,13.333); \node at (22.5,13.333) {$b=\lambda_{2}$};
		\draw[black,thick] (15,8)--(20,10.667); \node at (22.5,10.667) {$b=\lambda_{1}$}; \node at (22,15.75) {$\vdots$};

		\node at (1.5,4) {$\bphase{r}$};
		\node at (4.625,4) {$\cdots$};
		\node at (7.5,4) {$\bphase{j}$};
		\node at (10.625,4) {$\cdots$};
		\node at (13.5,4) {$\bphase{1}$};
		\node at (17.75,4) {${\phase{0}}$};
		
		\node at (17.75,10.5) {$\phase{1}$};
		\node at (15.75,12.25) {$\ddots$};
		\node at (12.75,14) {$\phase{j}$};
		\node at (8,14) {$\cdots$};
		\node at (1.5,14) {$\phase{r}$};
		\node at (0,-0.75) {0};\node at (21,0) {$\lambda$};\node at (0,21) {$b$};
		\node at (3,-0.75) {$A_{r}$}; 
		\node at (6.6,-0.85) {$A_{j+1}$};
		\node at (9,-0.85) {$A_{j}$};
		\node at (12,-0.75) {$A_{2}$};
		\node at (15,-0.75) {$A_{1}$};
		\node at (-0.75,0) {0};\node at (-0.75,8) {1};
		\node at (4.5,-0.75) {$\cdots$};
		\node at (10.5,-0.75) {$\cdots$};
		\end{tikzpicture}
		
		\caption{A slice of the phase diagram for the \olasep{} determined by parameters $(s_1,\dots,s_{r-1})$. The values of $\lambda_i$'s are determined by \eqref{lambdai-def} and the coordinates of the points $A_i$ are given by \eqref{Ai-coords}.}
		\label{fig:genrpd}
	\end{center}
\end{figure}
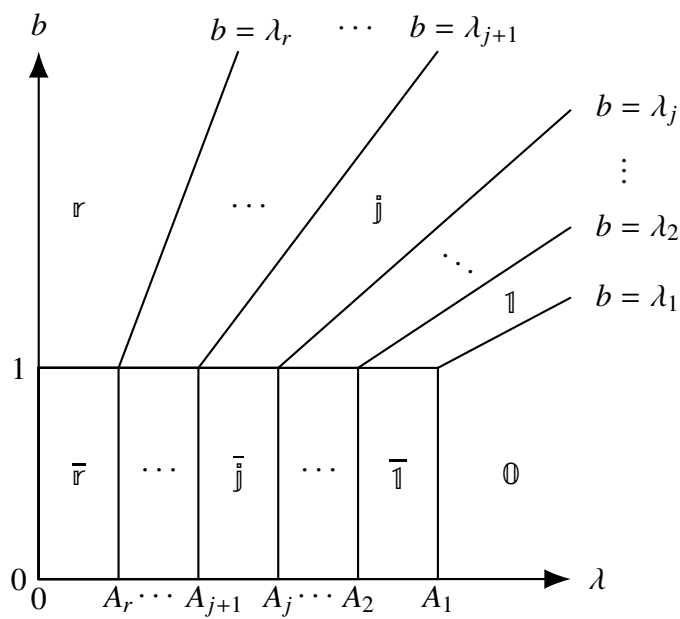

\subsection{Phase diagram}
\label{ssec:pdgenr}
We obtain the phase diagram of the \olasep{} by dividing the $(r+1)$-dimensional phase space appropriately. In order to find all phases, we consider the phase diagram for the model obtained by $k$-colouring for all $k$ simultaneously. The three phase regions in the LPASEP phase diagram in Figure~\ref{fig:lasepphdiag} lead to a total of $2r+1$ phase regions as follows.

\begin{itemize}
\item Phase $\bphase{r}$: $\max \left\lbrace \lambda_{r},b \right\rbrace <1$.

\item For $1 \leq j \leq r-1$, phase $\bphase{j} $:  $\max \left\lbrace \lambda_{j},b \right\rbrace <1<\lambda_{j+1}$.

\item Phase $\phase{0}$: $\max \left\lbrace 1,b \right\rbrace  < \lambda_{1}$.

\item For $1 \leq j \leq r-1$, phase $\phase{j}$: $\max \left\lbrace  1,\lambda_{j} \right\rbrace  < b < \lambda_{j+1}$.

\item Phase $\phase{r}$: $\max \left\lbrace  1,\lambda_{r} \right\rbrace  < b$.
\end{itemize}

In order to visualize the phases, we fix real constants $(s_1,\dots,s_{r-1})$ such that $s_1 > \dots > s_{r-1} > 1$. Consider the  two-dimensional plane determined by 
\begin{equation}
\label{lambdai-def}
\lambda_r = s_{r-1} \lambda_{r-1}= \cdots = s_{1} \lambda_{1}. 
\end{equation}
On this plane, $\lambda = \left( \sum_{i=1}^{r} \lambda_{i}^2 \right)^{1/2}$ measures the radial distance from origin in the $(\lambda_{1}, \ldots , \lambda_{r})$-subspace. 
Let $A_{i}$ be the point on this plane at which the hyperplanes $b=0$ and $\lambda_{i} =1$ intersect. Then $A_{i}$ has coordinates
given by
\begin{equation}
\label{Ai-coords}
\begin{cases}
(s_{i}/s_{1} , \ldots , s_{i}/s_{i-1}, 1, s_{i}/s_{i+1}, \ldots, s_{i}/s_{r-1}, s_{i}, 0) & 1 \leq i < r, \\
(1/s_{1},\ldots, 1/s_{r-1}, 1, 0) & i=r. 
\end{cases}
\end{equation}
We draw the two-dimensional phase diagram in terms of $A_{i}$'s and the parameters $\lambda$ and $b$. 
The phase regions are as illustrated in Figure~\ref{fig:genrpd}. 
The line $b=\lambda_{j}$ is the boundary between phases $\phase{j} -\phase{1}$ and $\phase{j}$ for $1 \leq j \leq r$. 
We now describe the currents and densities in each phase.

\begin{table}[htbp!]
	\renewcommand*{\arraystretch}{1.25}
	\begin{center}
		\begin{tabular}{|c|c|c|c|}
			\hline 
			Phase & Species  & Density $\rho$ & Current $J$  \\ \hline
			\multirow{3}{*}{$\bphase{r}$} & $\overline{r}$ & $ f(1)$ &\\ \cline{2-4}
			& $\overline{r}< i <r$ & $ 0$ & 0\\ \cline{2-4}
			& $r$& $ f(1)$ & $g(1, 0)$ \\
			\hline
			\multirow{7}{*}{$\bphase{j}$} & $\overline{r}$ & $  f(\lambda_{r})  $ & \\ \cline{2-4}
			& $ i<\overline{j} $ & $ f(\lambda_{|i|}) - f(\lambda_{|i|+1}) $ &  \\ \cline{2-4}
			& $\overline{j} $ & $ f(1) - f(\lambda_{j+1})$ & \\ \cline{2-4}
			& $\overline{j} <i <j$ & $ 0 $ & 0\\ \cline{2-4}
			& $j$ & $ f(1) - f(\lambda_{j+1})$ & $g(1, \lambda_{j+1}) $ \\ \cline{2-4}
			& $ i >j$& $ f(\lambda_{i}) - f(\lambda_{i+1})$& $g(\lambda_{i}, \lambda_{i+1}) $ \\ \cline{2-4}
			& $r$& $ f(\lambda_{r})$ & $g(\lambda_{r}, 0) $ \\
			\hline
			\multirow{5}{*}{$\phase{0}$}& $\overline{r}$ & $ f(\lambda_{r}) $ & \\ \cline{2-4} 
			& $ \overline{r} < i <  0$& $f(\lambda_{|i|}) - f(\lambda_{|i|+1})$ &  \\ \cline{2-4}
			& $0$ & $1-2f(\lambda_{1})$ & 0\\ \cline{2-4}
			& $ 0 < i <  r$& $f(\lambda_{i}) - f(\lambda_{i+1})$ & $g(\lambda_{i}, \lambda_{i+1}) $ \\ \cline{2-4}
			& $r$& $ f(\lambda_{r})$ & $g(\lambda_{r}, 0)$ \\ 
			\hline
			\multirow{7}{*}{$\phase{j}$} & $ \overline{r} $& $ f(\lambda_{r}) $&  \\ \cline{2-4}
			& $ i < \overline{j} $ & $ f(\lambda_{|i|}) - f(\lambda_{|i|+1}) $ & \\ \cline{2-4}
			& $\overline{j}$& $ f(b) - f(\lambda_{j+1})$ &  \\ \cline{2-4}
			& $ \overline{j} < i <j$ & $ 0 $ & 0\\ \cline{2-4}
			& $j$& $ \overline{f}(b) - f(\lambda_{j+1})$ & $g(b, \lambda_{j+1}) $ \\ \cline{2-4}
			& $ i >j$& $ f(\lambda_{i}) - f(\lambda_{i+1}) $& $g(\lambda_{i}, \lambda_{i+1}) $ \\ \cline{2-4}
			& $r$& $ f(\lambda_{r}) $& $g(\lambda_{r}, 0) $ \\
			\hline
			\multirow{3}{*}{$\phase{r}$} 
			& $\overline{r}$& $ f(b)$ & \\ \cline{2-4}
			& $ \overline{r} < i <  r$ & $ 0$ & 0 \\ \cline{2-4}
			& $r$& $ \overline{f}(b)$ & $g(b, 0)$ \\ 
			\hline
		\end{tabular}
	\end{center}
	\vspace{0.4cm}	
	\caption{Bulk densities and currents in each phase for \olasep{}. We do not write the currents for barred species since $J_{\overline{i} } = -J_{i}$.} 
	\label{table:cdmlasep}
\end{table}

\subsection{Currents and Densities}
We explain how to calculate densities and currents using $k$-colouring for the \olasep{}. We will give all the details only for phase $\phase{0}$ and sketch the argument for other phases. 
To describe the densities and currents succinctly, we define $f(x) = 1/(1+x)$, $\overline{f}(x) = 1- f(x),$ and $g(x,y) = (1-q) \left( f(x) \overline{f}(x) - f(y)\overline{f}(y) \right)$. The results are tabulated in Table~\ref{table:cdmlasep} and summarized below. 

Before we go on to the calculation, we make a couple of remarks about the currents. The currents of barred species are determined completely by their unbarred partners; specifically $J_{\overline{i}} = -J_{i}$. This is because every species $i$ can enter and exit the right boundary only at the expense of its barred partner. Moreover, since $0$'s can neither enter or leave from the right boundary, there is no current of species 0, i.e. $J_{0}=0$.

\subsubsection*{ Phase $\phase{0}$}
In phase $\phase{0}$, the \olasep{} is projected onto the LD phase of the LPASEP by all colourings. Hence, we have from Table~\ref{table:cdlpasep}
\[
\rho_{0} = 1 -2 f(\lambda_{1}), \quad \sum_{i=1}^{r} \rho_{i} = \sum_{i=\overline{r}}^{\overline{1}} \rho_{i} = f(\lambda_{1}),
\]
and $\sum_{i=1}^{r} J_{i} = g(\lambda_{1},0)$ by the 1-colouring. Similarly, $ 2 \leq k \leq r$, one obtains 
\[
\sum_{i=\overline{k-1}}^{k-1} \rho_{i} = 1 -2 f(\lambda_{k}), \quad
\sum_{i=k}^{r} \rho_{i} = \sum_{i=\overline{r}}^{\overline{k}} \rho_{i} = f(\lambda_{k})
\]
and $\sum_{i=k}^{r} J_{i} = g(\lambda_{k},0)$ by $k$-colouring. 
Comparing the $k$ and $(k+1)$-colouring, we find that $\rho_{k} = \rho_{\overline{k}} = f(\lambda_{k}) - f(\lambda_{k+1})$ and $ J_{k}= g(\lambda_{k},\lambda_{k+1})$ for $1 \leq k < r$. Finally, by the $r$-colouring, one obtains $\rho_{r} = \rho_{\overline{r}} = f(\lambda_{r})$, and $J_{r}= g(\lambda_{r},0)$.
See Figure~\ref{fig:Asimd}(c) for the densities in phase $\phase{0}$ of the \olasep{} with $r=2$.

\subsubsection*{ Phases $\phase{j}$ and $\bphase{j}$ }
Here, the $k$-colouring maps phase $\phase{j}$ to the LD (resp. HD) phase of the LPASEP and phase $\bphase{j}$ to the LD (resp. MC) phase of an LPASEP for $k\geq j$ (resp. $k<j$).
In these phases, we have (i) $\rho_{i} = \rho_{\overline{i}}$ for all $i>j$, and (ii) all species $i$ with $\overline{j}<i<j$ are dynamically expelled, i.e. $\rho_{i}=0$. 
See Figures~\ref{fig:Asimd}(b) and (d) for the densities in phases $\bphase{1}$ and $\phase 1$ of the \olasep{} with $r=2$.

\subsubsection*{ Phases $\phase{r}$ and $\bphase{r}$} 
All $k$-colourings now map phase $\phase{r}$ and $\bphase{r}$ onto HD and MC phases respectively of the LPASEP. Thus all species $i$ satisfying $\overline{r}<i<r$ are dynamically expelled. The density and current of species $r$ are then given by the $r$-colouring.
See Figures~\ref{fig:Asimd}(a) and (e) for the densities in phases $\bphase{2}$ and $\phase 2$ of the \olasep{} with $r=2$.

\subsubsection*{$(\phase{j}-\phase{1})-\phase{j}$ co-existence line} 
\begin{figure}[htbp!]
	
	\includegraphics[width=1\linewidth]{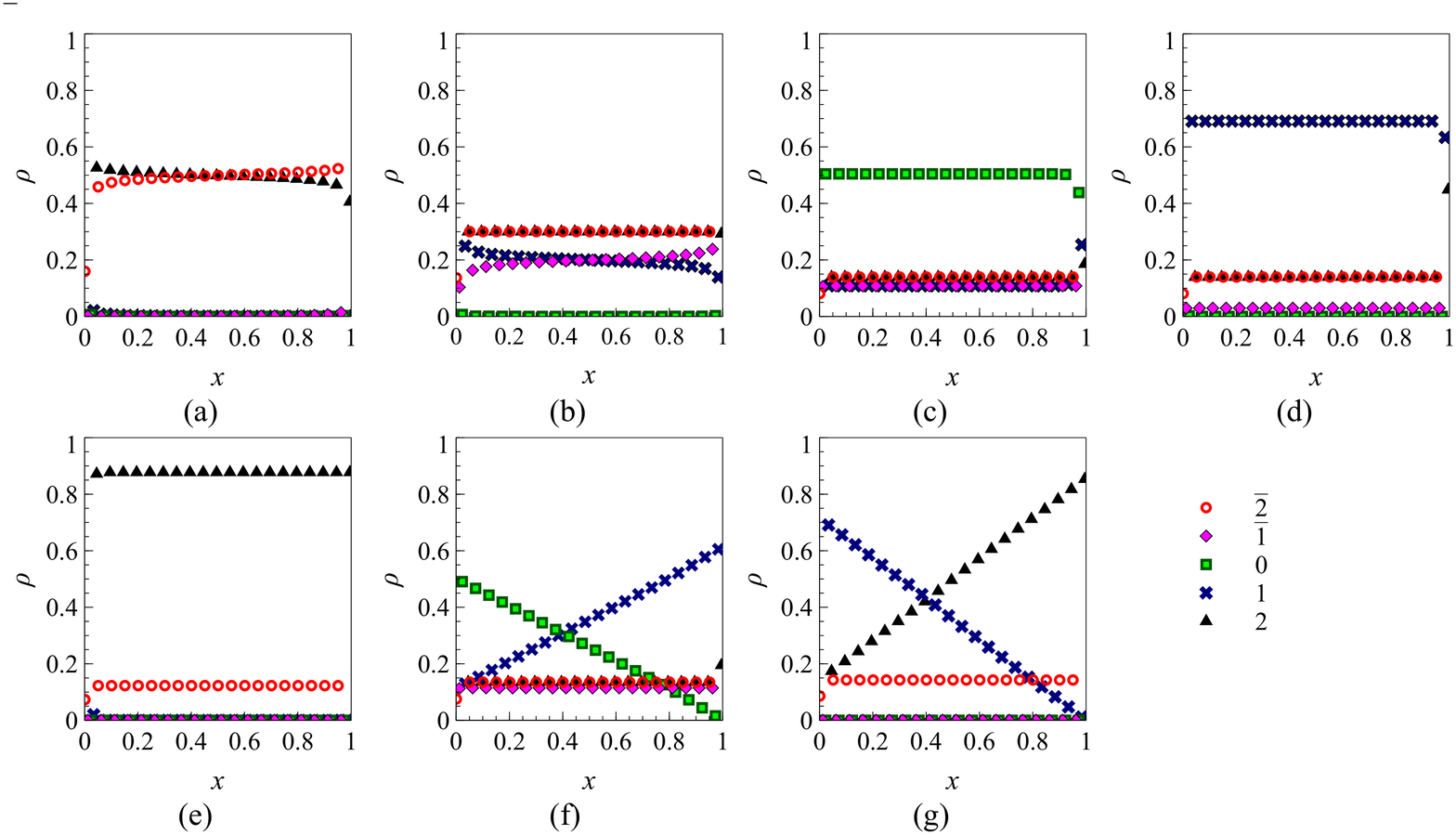}
	
	\caption{Time-average densities in 5-species \olasep{} for species $\overline{2}$ (red circles), $\overline{1}$ (magenta diamonds), 0 (green boxes), 1 (blue crosses) and 2 (black triangles) for 
		(a) phase $\bphase{2}$ ($ \lambda_1 \simeq 0.36 ,\lambda_2 \simeq 0.8, b \simeq 0.59$),
		(b) phase $\bphase{1}$ ($ \lambda_1 \simeq 0.37 ,\lambda_2 \simeq 2.33, b \simeq 0.59$), 
		(c) phase $\phase{0}$ ($ \lambda_1 \simeq 3.04 ,\lambda_2 \simeq 6.16, b \simeq 2.61$),  
		(d) phase $\phase{1}$ ($ \lambda_1 \simeq 3.04 ,\lambda_2 \simeq 6.16, b \simeq 4.91$), 
		(e) phase $\phase{2}$ ($ \lambda_1 \simeq 3.02,\lambda_2 \simeq 6.09, b \simeq 7.15$), 
		(f) $\phase{0}-\phase{1}$ coexistence line ($ \lambda_1 = 3 ,\lambda_2 \simeq 6.41, b = 3 $), 
		and (g) $\vvmathbb{1}-\vvmathbb{2}$ coexistence line ($ \lambda_1 \simeq 3.05 ,\lambda_2 = 6, b =6$). The lattice size is $1000$ for all cases. }
	\label{fig:Asimd}
\end{figure}

The $k$-colouring maps the $(\phase{j}-\phase{1})-\phase{j}$ boundary to the HD-LD co-existence line of the LPASEP for $k=j$, and to the LD (resp. HD) phase of the LPASEP for $k> j$ (resp. $k<j$). All species $i$ with $\overline{j}<i<j-1$ are dynamically expelled. Moreover, species $j-1$ and $j$ have linear densities on these lines.
See Figures~\ref{fig:Asimd}(f) and (g) for the densities on the $\phase{1}-\phase{2}$ and $\phase{0}-\phase{1}$ coexistence lines of the \olasep{} with $r=2$.

\subsection{Example of $r=2$}
\label{sec:pd-r=2}

The simplest nontrivial \olasep{} is the one with five species. The boundary transitions are given by
\begin{center}
	\begin{tabular}{ll}
		\underline{Left}:
		&$\begin{cases}
		\overline{2},\overline{1}, 0,1 \rightarrow 2 \quad \text{ with rate } \alpha_{2},	\\
		\overline{2},\overline{1}, 0 \rightarrow 1 \quad \text{ with rate } \alpha_{1},	\\
		\overline{2},\overline{1} \rightarrow 0 \quad \text{ with rate } \alpha_{0},	\\
		1,2 \rightarrow 0 \quad \text{ with rate } \gamma_{0}, \\
		2 \rightarrow 1 \quad \text{ with rate } \gamma_{1}, 
		\end{cases}$ \\
		&\\
		\underline{Right}:
		&$\begin{cases}
		2 \rightarrow \overline{2} \quad \text{ with rate } \beta, \\
		1 \rightarrow \overline{1} \quad \text{ with rate } \beta, \\
		\overline{1} \rightarrow 1 \quad \text{ with rate } \delta, \\
		\overline{2} \rightarrow 2 \quad \text{ with rate } \delta. 
		
		\end{cases}$
	\end{tabular}
\end{center}

 In the 1-colouring, we identify species 1's and 2's as $1_{1}$, $\overline{1}$'s and $\overline{2}$'s as $\overline{1}_{1}$, and 0's as $0_{1}$, such that the rates for boundary transitions are given by
 \begin{center}
 	\begin{tabular}{ll}
		\underline{Left}:
 		&$ \begin{cases}
 		\overline{1}_{1}, 0_{1} \rightarrow 1_{1} \quad \text{ with rate } \theta_{1},	\\
 		\overline{1}_{1} \rightarrow 0_{1} \quad \text{ with rate } \phi_{1}, \\
 		1_{1} \rightarrow 0_{1} \quad \text{ with rate } \zeta_{1} ,
 		\end{cases}$ \\
 		&\\
 		\underline{Right}:
 		&$\begin{cases}
 		1_{1} \rightarrow \overline{1}_{1} \quad \text{ with rate } \beta, \\
 		\overline{1}_{1} \rightarrow 1_{1} \quad \text{ with rate } \delta.
 		\end{cases}$
 	\end{tabular}
 \end{center}
 The relevant left and right boundary parameters are $\lambda_{1}$ and $b= \kappa_{\beta, \delta}^{+}$ respectively. On the other hand, we label $\overline{1}$'s,0's and 1's with $0_{2}$, $2$'s with $1_{2}$, and $\overline{2}$'s with $\overline{1}_{2}$ in 2-colouring. Now, we have the following boundary rates 
 \begin{center}
 	\begin{tabular}{ll}
 		\underline{Left}:
 		&$\begin{cases}
 		\overline{1}_{2}, 0_{2} \rightarrow 1_{2} \quad \text{ with rate } \theta_{2},	\\
 		\overline{1}_{2} \rightarrow 0_{2} \quad \text{ with rate } \phi_{2}, \\
 		1_{2} \rightarrow 0_{2} \quad \text{ with rate } \zeta_{2} , 
 		\end{cases}$ \\
 		&\\
 		\underline{Right}:
 		&$\begin{cases}
 		1_{2} \rightarrow \overline{1}_{2} \quad \text{ with rate } \beta, \\
 		\overline{1}_{2} \rightarrow 1_{2} \quad \text{ with rate } \delta.
 		\end{cases}$
 	\end{tabular}
 \end{center}
The relevant parameters $\lambda_{2}$ and $b= \kappa_{\beta, \delta}^{+}$ correspond to the left and right boundary respectively.

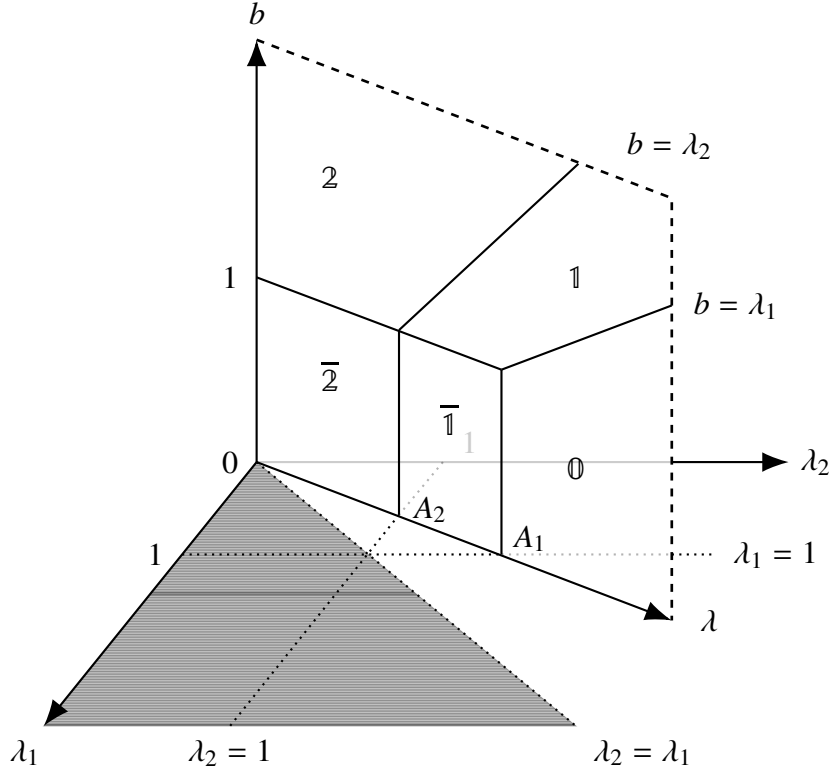
\begin{figure}[htbp!]
	\begin{tikzpicture}[scale=0.35]
	\draw[-{Latex[length=3.5mm]},thick] (0,0)--(0,16);
	\draw[black,line width = 0.8pt, opacity =0.2] (0,0)--(15.7,0);
	\draw[-{Latex[length=3.5mm]},thick] (15.62,0)--(20,0);
	\draw[-{Latex[length=3.5mm]},thick] (0,0)--(-8,-10);
	\draw[black, thick, dotted] (0,0) -- (12,-10);
	\draw[-{Latex[length=3.5mm]}, thick] (0,0)--(15.6,-6);
	\draw[black, thick, dotted, opacity=0.3] (7,0)--(5.35,-2);
	\draw[black, thick, dotted] (5.35,-2)--(-1,-10); \node at (-1,-11) {$\lambda_{2}=1$};
	\draw[black, line width = 1pt, dashed] (0,16)--(15.6,10)--(15.6,-6);
	\draw[black, thick, dotted] (-2.8,-3.5)--(9.2,-3.5);\draw[black, thick, dotted] (15.6,-3.5)--(17.2,-3.5);
	\draw[black, thick, dotted, opacity = 0.3] (9.2,-3.5)--(15.6,-3.5);
	\draw[black,thick] (0,7)--(9.2,3.5);
	\draw[black,thick] (9.2,-3.5)--(9.2,3.5);
	\draw[black,thick] (5.35,-2)--(5.35,5);
	\draw[black,thick] (12.091,11.3)--(5.35,5); 
	\node at (15.5,12) {$b= \lambda_{2}$};
	\draw[black,thick] (15.6,5.9348)--(9.2,3.5); 
	\node at (18,5.9348) {$b= \lambda_{1}$};
	
	\draw[top color=black, bottom color=black, opacity= 0.25] (0,0) -- (-8,-10) -- (12,-10) -- cycle; 
	
	\node at (0,17) {$b$}; \node at (-1,7) {$1$}; \node at (-1, 0) {0};
	\node at (-3.8, -3.5) {1}; \node at (19.5, -3.5) {$\lambda_{1} = 1$};
	\node at (17,-6) {$\lambda$};
	\node at (21,0) {$\lambda_{2}$}; \node[opacity =0.2 ] at (8,0.75) {$1$};
	\node at (-8.7,-11) {$\lambda_{1}$}; 
	\node at (14.5,-11) {$\lambda_{2} = \lambda_{1}$};
	
	\node at (6.5,-1.8) {$A_{2}$};
	\node at (10.25,-2.9) {$A_{1}$};
	
	\node at (2.75, 3.25) {$\bphase{2}$};
	\node at (7.25, 1.6) {$\bphase{1}$};
	\node at (12, -0.25) {$\phase{0}$};
	\node at (12, 7) {$\phase{1}$};
	\node at (2.75, 10.8) {$\phase{2}$};
	\end{tikzpicture}
	
	\caption{Phase diagram for mLPASEP with $r=2$. The shaded region ($\lambda_{1} > \lambda_{2}$) is forbidden. The plane in focus is given by $\lambda_2 = s_1 \lambda_1$.}
	\label{fig:2pd}
\end{figure}

The phase diagram is the three-dimensional space of the parameters $\lambda_{1}, \lambda_{2}$ and $b$. The region $\lambda_{1} > \lambda_{2}$ is excluded. We fix a constant $s_{1}>1$ and consider the two dimensional plane $\lambda_{2}= s_{1} \lambda_{1}$. 
In this plane, $\lambda$ is the distance along the $b=0$ plane.
This plane passes through all the phases and, as a result, allows us to visualize all phases on a two-dimensional phase diagram as shown in Figure~\ref{fig:2pd}. Using the colouring ideas as outlined in Section~\ref{sec:genpd}, we find that there are five phases in the phase diagram: 

\begin{itemize}

\item phase $\bphase{2}$: $ \max\left\lbrace b, \lambda_{2} \right\rbrace < 1 $,

\item phase $\bphase{1}$: $\max\left\lbrace b, \lambda_{1}\right\rbrace  < 1 < \lambda_{2}$, 

\item  phase $\phase{0}$: $\max\left\lbrace 1,b \right\rbrace <\lambda_{1}$,

\item phase $\phase{1}$: $ \max \left\lbrace 1, \lambda_{1} \right\rbrace < b < \lambda_{2}$,

\item phase $\phase{2}$: $\max\left\lbrace 1, \lambda_{2}\right\rbrace  <b$.

\end{itemize}

In addition, the two co-existence planes, the $\phase{0} - \phase{1}$ phase boundary: $1< b = \lambda_{1}$ , and the $\phase{1} - \phase{2}$ phase boundary: $1< b = \lambda_{2}$, appear as lines.
On the $b=0$ plane, the plane $\lambda_{i}=1$ and the plane $\lambda_{2} = s_{1} \lambda_{1}$ intersect at the point denoted by $A_{i}$ for $i=1,2$. $A_{1}$ and $A_{2}$ have locations $(1,s_{1},0)$ and $(1/s_{1},1,0)$.

\section{The shock picture in the \olasep{}} 
\label{sec:gshockpicture}
We now use the shock picture in the LPASEP to understand the density profiles as well as the phenomenon of dynamical expulsion in the \olasep{}. We will explain this picture in each phase and phase-boundary.
This picture is best understood by looking at the coexistence lines first.

\begin{figure}[htbp!]
	\begin{center} 
		\begin{tikzpicture}[scale=0.75]
		\draw[black,thick] (0,0) rectangle (10,10);
		\draw[black] (0,4)--(6,4)--(6,6);
		
		\draw[black] (0,3)--(10,3);
		\draw[black] (0,6)--(10,6);

		\draw[black] (0,1)--(10,1);
		\draw[black] (0,2)--(10,2);
		\draw[black] (0,7)--(10,7);
		\draw[black] (0,8)--(10,8);
		\draw[black] (0,9)--(10,9);
		
		\node[] at (-1,4){$\scriptstyle f(\lambda_{j})$};
		\node[] at (-1.2,3){$\scriptstyle f(\lambda_{j+1})$};
		\node[] at (-1,1){$\scriptstyle f(\lambda_{r})$};
		\node[] at (11.2,7){$\scriptstyle\overline{f}(\lambda_{j+1})$};
		\node[] at (11,6){$\scriptstyle\overline{f}(\lambda_{j})$};
		\node[] at (11,9){$\scriptstyle\overline{f}(\lambda_{r})$};
		\node[] at (-1,2){$\vdots$};
		\node[] at (11,8){$\vdots$};
		
		\node[] at (5,7.5){$\overline{j+1}$}; 
		\node[] at (5,6.5){$\overline{j}$}; 
		\node[] at (5,9.5){$\overline{r}$}; 
		\node[] at (5,0.5){$r$};
		\node[] at (5,2.5){$j+1$};
		\node[] at (5,8.6){$\vdots$};
		\node[] at (5,1.6){$\vdots$};
		\node[] at (3,5){$j-1$};
		\node[] at (7.5,5){$j$};
		\draw[<->] (5.5,5)--(6.5,5);
		
		\node at (-1.8,5) {$\rho$};
		\node[] at (0,-0.5){$0$};\node[] at (10,-0.5){$1$};
		\node[] at (-0.5,0){$0$}; \node[] at (10.57,10){1};
		\node[] at (5,-1){$x$};
		
		\end{tikzpicture}
		
	\end{center}
	\caption{The shock picture for mLPASEP: The schematic plot shows the shock on the co-existence line $b=\lambda_{j}>1$ for the \olasep{} where $1\leq j \leq r$. At each point $x$, the densities of the $k$-coloured species $1_k$ and $\overline{1}_k$ satisfy \eqref{dens-pairs} for each $k$.}
	\label{fig:gshockr}
\end{figure}
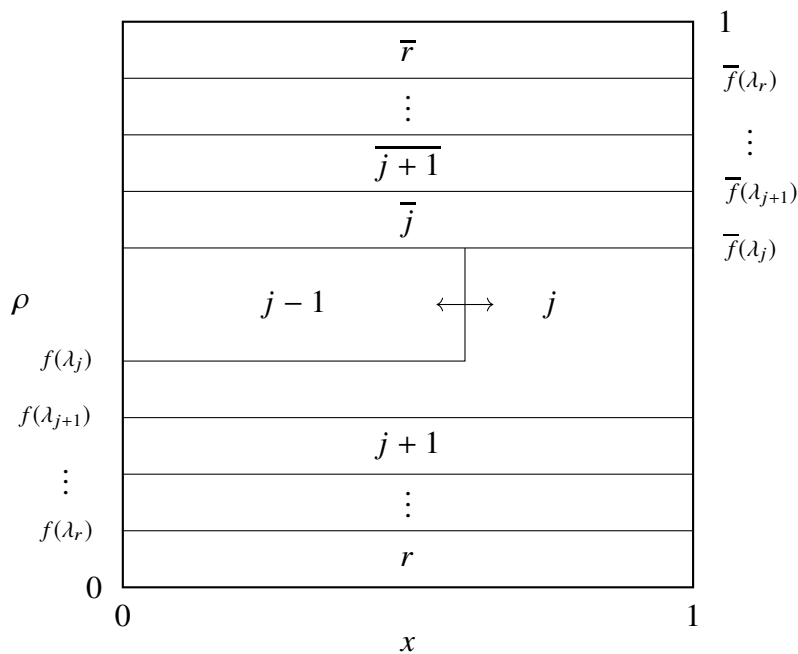

\begin{figure}[htbp!]
	\includegraphics[width=0.8 \linewidth]{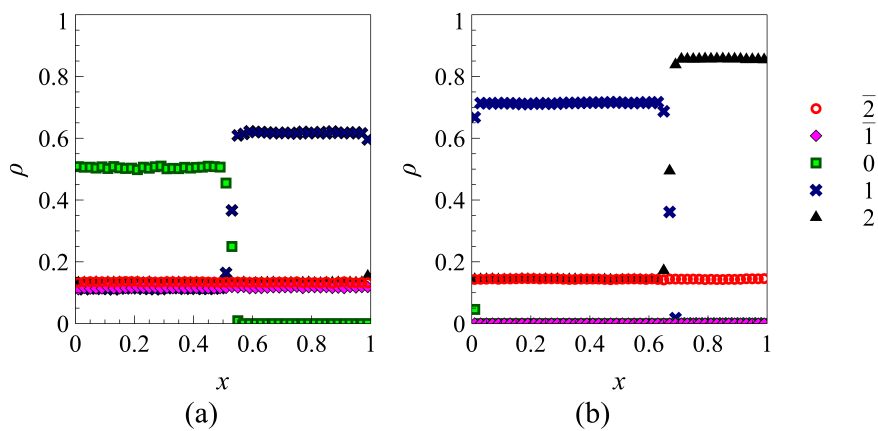}
	\caption{Instantaneous shock profiles in 5-species \olasep{} : Density profiles for species $\overline{2}$ (red circles), $\overline{1}$ (magenta diamonds), 0 (green boxes), 1 (blue crosses) and 2 (black triangles) on (a) $\phase{0}-\phase{1}$ coexistence line ($ \lambda_1 = 3 ,\lambda_2 \simeq 6.41, b = 3 $), and (b) $\vvmathbb{1}-\vvmathbb{2}$ coexistence line ($\lambda_1 \simeq 3.05 ,\lambda_2 = 6, b =6$), where the lattice size is $2500$.  }
	\label{fig:5asepcg}
\end{figure}

\noindent
\textbf{The $ (\phase{j}-\phase{1}) - \phase{j} $ coexistence line for $1 \leq j \leq r$}:
From the $k$-colouring argument, we see that species $j-1$ and $j$ are phase-segregated and other species have constant densities on this coexistence line. In other words, only species $j-1$ and $j$ take part in the shock on this coexistence line. This is illustrated in the schematic plot in Figure~\ref{fig:gshockr}. The shock performs a random walk with no net drift. Moreover, species $\overline{j-1}, \ldots , j-2$ are dynamically expelled. 
See Figure~\ref{fig:Asimd}(f) and (g) for simulations of the \olasep{} with $r=2$ on the $\phase{0} - \phase{1}$ and $\phase{1} - \phase{2}$ coexistence lines respectively. See also Figures~\ref{fig:5asepcg} (a) and (b) for instantaneous profiles of the shock in these lines.

\noindent
\textbf{Phases $\bphase{j}$ and $\phase{j}$ for $1 \leq j \leq r$}:
In phase $\phase{j}$, the shock front is pinned to the left causing the dynamical expulsion of species $j-1$ and higher density of $j$'s compared to $\overline{j}$'s.
In phase $\bphase{j}$, the $j-1$'s are again dynamically expelled because the height of the shock vanishes.
See Figure~\ref{fig:Asimd}(a), (b), (d) and (e) for simulations of the \olasep{} with $r=2$ in phases $\bphase{2}, \bphase{1}, \phase 1$ and $\phase 2$.

\noindent
\textbf{{Phase $\phase{0}$}}:
The shock picture on the $\phase{0}-\phase{1}$ coexistence line is Figure~\ref{fig:gshockr} with $j=1$. 
The shock front has positive drift in phase $\phase{0}$ and consequently gets pinned to the right boundary resulting in non-zero bulk density of species 0. Hence, all species have non-zero densities in phase $\phase{0}$.  
See Figure~\ref{fig:Asimd}(c) for simulations of the \olasep with $r=2$ in phase $\phase{0}$.

\begin{figure}[htbp!]
	\begin{tabular}{ *2{c} }
		\includegraphics[width=.45\linewidth]{{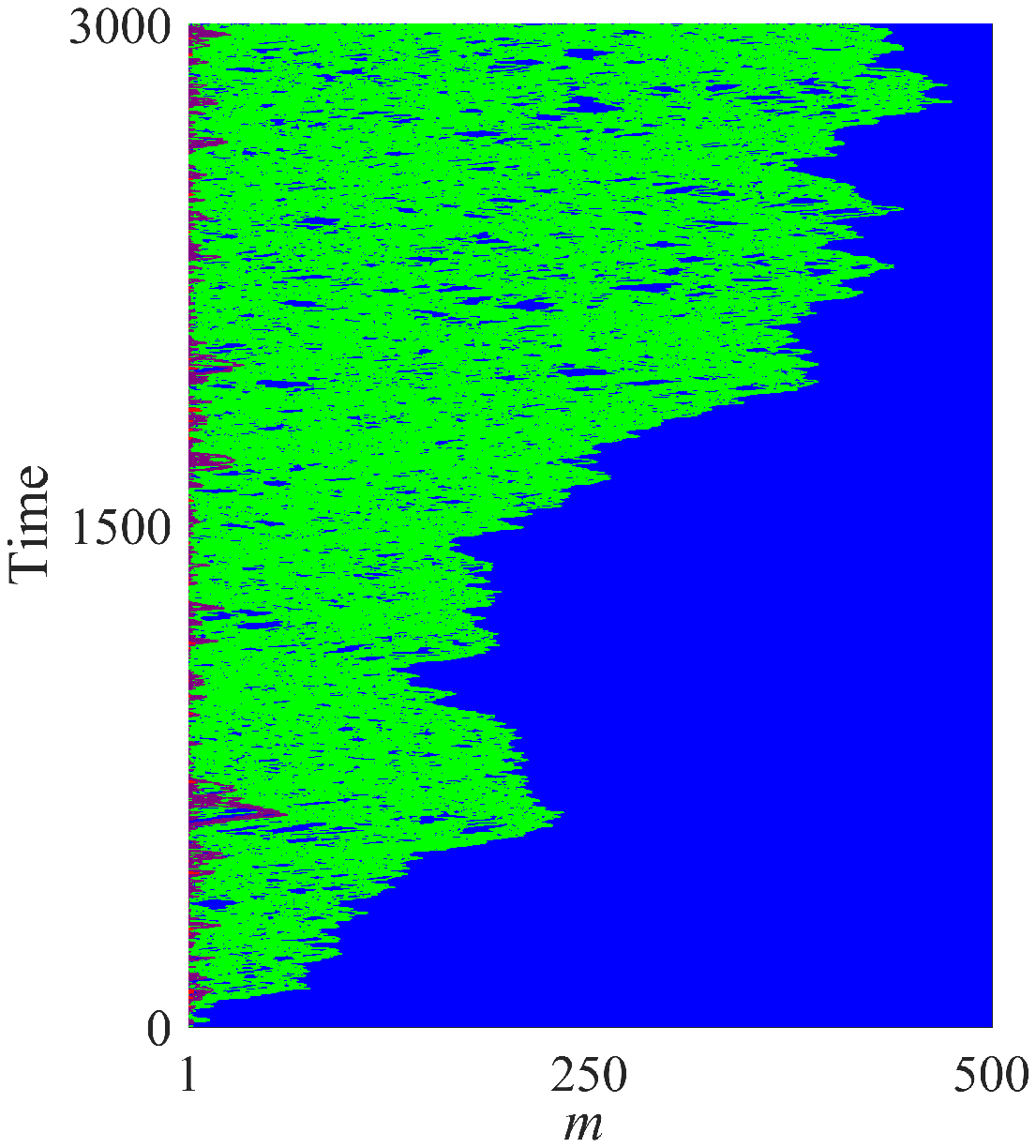}} & 
		\includegraphics[width=.45\linewidth]{{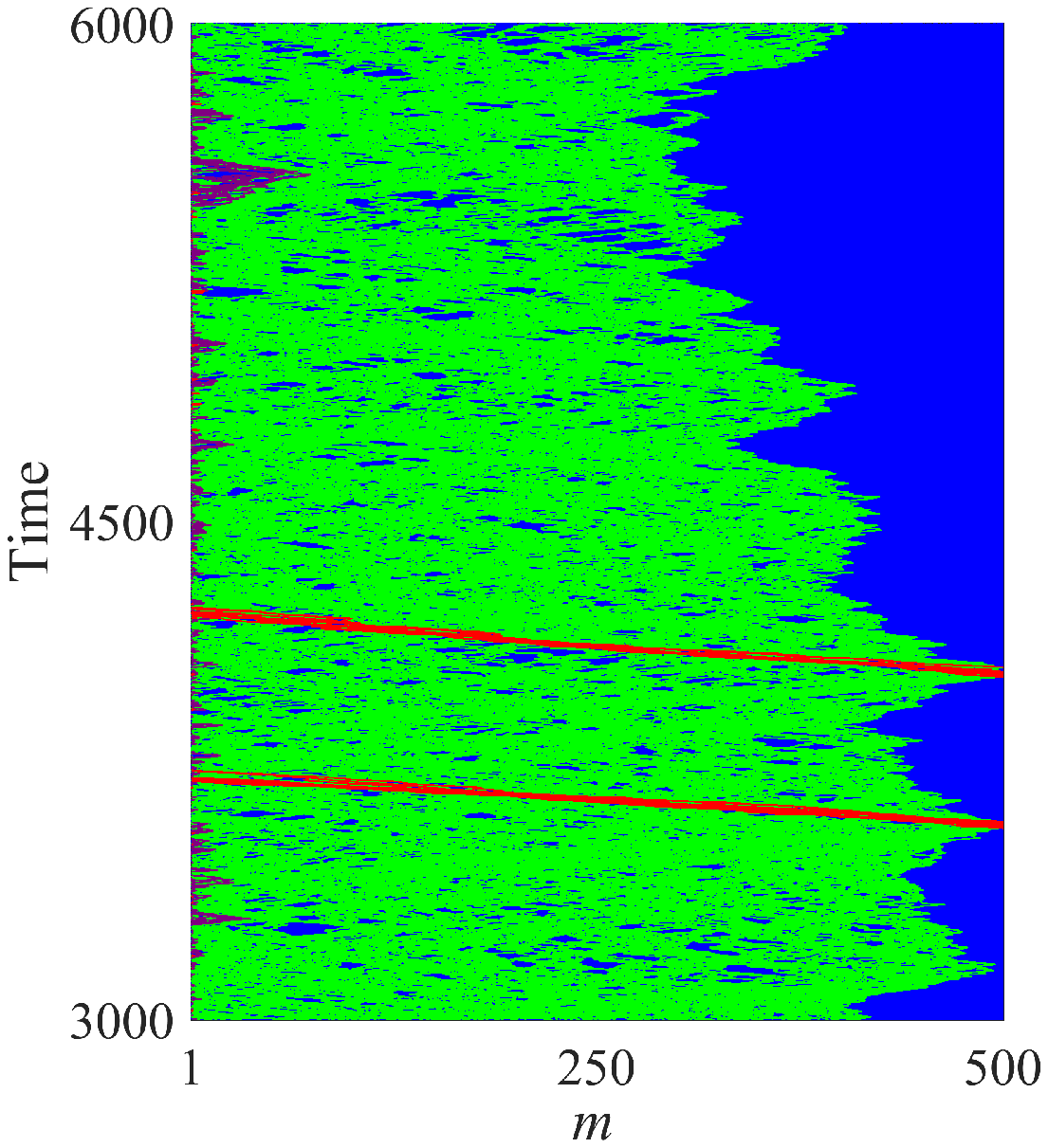}} \\ 
		(a)  & (b) \\
	\end{tabular}
	\caption{
	The spatio-temporal evolution of the shock on the $\phase{1}-\phase{2}$ co-existence line for 5-species \olasep{} for 6000 time steps. Each time step equals 4000 random sequential updates in the simulation.
The plots show trajectories of species $\overline{1}$ (red), 0 (violet), 1 (green) and 2 (blue) versus site position after the system reaches steady state.  Particles of species $\overline 2$ are not shown.
 The parameters are as follows: $\alpha_{\overline{1}}=0.08, \alpha_{0}= 0.45, \alpha_{1} =0.13, \alpha_{2} =0.33, \gamma_{0} \simeq 0.005, \gamma_{1}\simeq 0.06, q=0.1, \beta = 0.475, \ \text{and} \ \delta = 0.35 \, (b=\lambda_{2} = 2, \lambda_{1} \simeq 0.98)$, and the lattice size is $500$.}
	\label{fig:5asepshocksim}
\end{figure}

In addition, we note the following on the $ (\phase{j}-\phase{1}) - \phase{j} $ coexistence line. Species $\overline{j-1}$ is dynamically expelled on this line, although species $j-1$ has non-zero bulk density. 
This might seem counterintuitive because \eqref{dens-pairs} suggests that either a species and its barred partner are both present or both absent. The resolution of this apparent contradiction is the fact that \eqref{dens-pairs} only applies to the $k$-coloured species
$\overline{1}_k$ and $1_k$ for each $k$.

To illustrate this point further, we perform a spatio-temporal simulation of the \olasep{} with $r=2$ on the 
$\phase{1}-\phase{2}$ coexistence line. 
The results of the simulation are shown in Figure~\ref{fig:5asepshocksim}.
The shock there is formed between species 1 and 2 and has zero mean velocity. Species $0$ and $\overline{1}$ are dynamically expelled. 
As one can see from the simulation, particles of species $\overline{1}$ can enter either on the left or the right boundary, but they eventually leave from the left boundary 
because of the high density of 1's and 2's. They can only enter at the right boundary when the $1-2$ shock touches the right boundary.

\section{Conclusion}
In this article, we have defined a multispecies ASEP and determined the exact phase diagram corresponding to the model. The structure of the phase diagram derived by the colouring method is the same when there are either $2r$ or $(2r+1)$-species in the model. 
It would be an interesting problem to find a matrix ansatz for the mLPASEP so that the densities and currents can be computed directly. 
The phase diagram of LPASEP has a rich structure that manifests itself in the presence of subphases inside the LD and HD phases. One might turn to the colouring technique to unearth the subphases in the phase diagram for the mLPASEP.

\section*{Acknowledgements}
We thank the referees for a number of useful suggestions.
The first and third authors are supported by UGC Centre for Advanced Studies. The first author was also partly supported by Department of Science and Technology grant EMR/2016/006624.

\appendix
\section{The \elasep{}}
\label{sec:ELASEP}
We explain the salient features of the phase diagram of the \elasep{} focusing on the aspects that make the analysis more complicated than that for the \olasep{}.

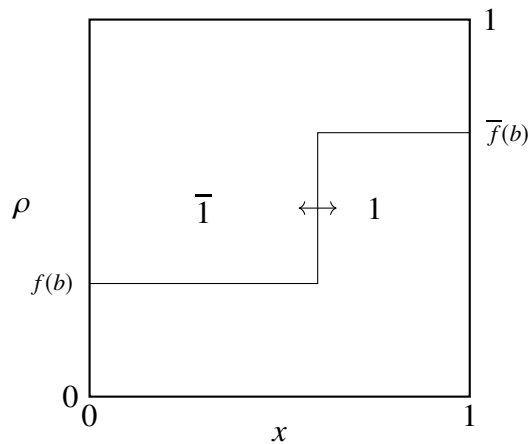
\begin{figure}[htbp!]
	\begin{center}
 	\begin{tikzpicture}[scale=0.5]
 			\draw[black,thick] (0,0) rectangle (10,10);
 			\draw[black] (0,3)--(6,3)--(6,7)--(10,7);
 			
 			\node[] at (-1,3){$\scriptstyle f(b)$};
 			
 			\node[] at (11,7){$\scriptstyle \overline{f}(b)$}; 
 			
 			\node[] at (3,5){$\overline{1}$};\node[] at (7.5,5){$1$};
 			\draw[<->] (5.5,5)--(6.5,5);
 			
 			\node at (-1.8,5) {$\rho$};
 			\node[] at (0,-0.5){$0$};\node[] at (10,-0.5){$1$};
 			\node[] at (-0.5,0){$0$}; \node[] at (10.55,10){1};
 			\node[] at (5,-1){$x$};
 	\end{tikzpicture}
 	\end{center}
 	
 	\caption{The shock picture for ASEP on the co-existence line $b=\lambda >1$.}
 	\label{fig:asepshock}
 \end{figure}

The computation of the generalized phase diagram for the \elasep{} requires us to project to the single-species ASEP \cite{uchiyama2004} which we review briefly. The ASEP involves only particles and vacancies, denoted by 1 and $\overline{1}$ respectively. The boundary transitions, in accord with our definitions in Eq.~\eqref{eq:elb} and \eqref{eq:rb}, have the following rates:
\begin{center}
	\begin{tabular}{ll}
		\underline{Left}:
		&$\begin{cases}
		\overline{1} \rightarrow 1 \quad \text{ with rate } \alpha_{1},	\\
		1 \rightarrow \overline{1}  \quad \text{ with rate } \gamma'_{0}, 
		\end{cases}$ \\
		&\\
		\underline{Right}:
		&$\begin{cases}
		1 \rightarrow \overline{1} \quad \text{ with rate } \beta , \\
		\overline{1} \rightarrow 1 \quad \text{ with rate } \delta.
		\end{cases}$
	\end{tabular}
\end{center}
The relevant left and right boundary parameters are $\lambda = \kappa_{\alpha_{1}, \gamma'_{0}}^{+}$ and $b = \kappa_{\beta, \delta}^{+}$ respectively, where $\kappa^\pm_{u,v}$ is defined in \eqref{def-kappa}. With this notation, the phase diagram 
formally looks exactly like Figure~\ref{fig:lasepphdiag} with the same nomenclature for the phases: phase $\bphase{1}$ (MC), $\phase{0}$ (LD) and $\phase{1}$ (HD). 
The currents and densities of species $1$ for all three phases 
are also identical to those in Table~\ref{table:cdlpasep}.
The density profiles for ASEP can also be understood by appealing to shocks. Since this is reviewed by Blythe and Evans in \cite{blythe2007}, we only illustrate the shock picture for the coexistence line $b=\lambda>1$ in Figure~\ref{fig:asepshock}, where the shock has zero mean velocity. In the LD (resp. HD) phase, the shock has positive (resp. negative) velocity and is pinned to the right (resp. left).

The phase diagrams for the \elasep{} with $2r$ species and the 
\olasep{} with $(2r+1)$ species have identical structure as depicted in Figure~\ref{fig:genrpd}. 
The main difference between the two is that the $1$-colouring projects the \elasep{} to the ASEP so that the boundary parameters are $\lambda_{1}=\kappa^{+}_{\theta_{1}, \gamma'_{0}}$ and $b=\kappa^{+}_{\beta, \delta}$. 
All other $k$-colourings continue to project the \elasep{} to the LPASEP with boundary parameters $\lambda_{k} = \theta_{k}/\phi'_{k}$ and $b= \kappa^{+}_{\beta, \delta}$., where $\theta_{k} = \sum_{i=k}^{r} \alpha_{i}$ and $\phi'_{i} = \sum_{i=1}^{k-1}\left(  \alpha_{i} +  \alpha_{\overline{i}} \right)$ were defined in Section~\ref{sec:defB}. 

\begin{figure}[htbp!]
	
	\includegraphics[width = 1 \linewidth]{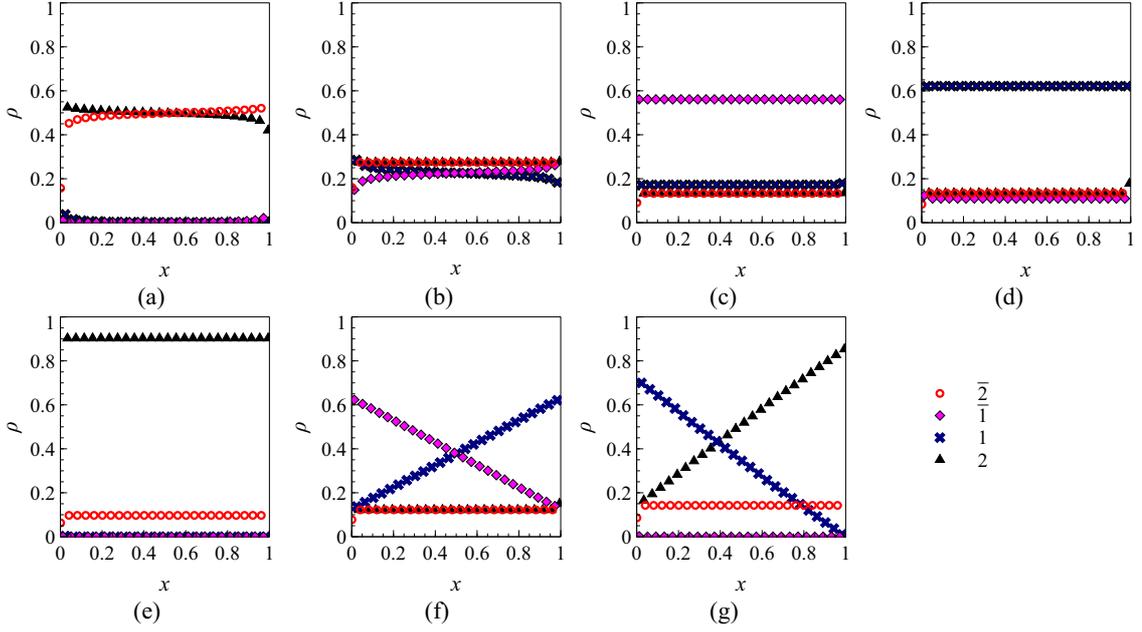}
	
	\caption{Time-average densities in 4-species mLPASEP for species $\overline{2}$ (red circles), $\overline{1}$ (magenta diamonds), 0 (green boxes), 1 (blue crosses) and 2 (black triangles) for 
		(a) phase $\bphase{2}$ ($ \lambda_1 \simeq 0.23 ,\lambda_2 \simeq 0.84, b \simeq 0.65$),	(b) phase $\bphase{1}$ ($ \lambda_1 \simeq 0.44, \lambda_2 \simeq 2.65, b \simeq 0.75$), (c) phase $0$ ($ \lambda_1 \simeq 2.27, \lambda_2 \simeq 6.5, b \simeq 0.92$), (d) phase $\vvmathbb{1}$ ($ \lambda_1 \simeq 1.17, \lambda_2 \simeq 6.45, b \simeq 3.09$), (e) phase $\vvmathbb{2}$ ($ \gamma_1 \simeq 0.12 , q=0.41, \beta=0.15,\delta=0.86, \lambda_1 \simeq 1.17, \lambda_2 \simeq 6.45, b \simeq 9.28$), (f) $\phase{0}-\phase{1}$ co-existence line \label{Bcl1} ($ \lambda_1 = 3, \lambda_2 \simeq 7.18, b = 3 $), and (g) $\vvmathbb{1}-\vvmathbb{2}$ co-existence line \label{Bcl2} ($ \lambda_1 \simeq 1.34, \lambda_2 = 6, b = 6$). For all simulations, we fix the lattice size to be $1000$.}
	\label{fig:Bsimd} 
\end{figure}

Taking into account all possible colourings, there are $r+1$ relevant boundary parameters, namely, $\lambda_{1}, \ldots, \lambda_{r}$ and $b$. Again, the inequalities $\lambda_{1} < \lambda_{2}< \ldots < \lambda_{r}$ are satisfied. Because of these relations among $\lambda_{i}$'s, we arrive at the same phase diagram in Figure~\ref{fig:genrpd} which shows all $2r+1$ phases in the \elasep{}. 
In all phases except phase $\phase{0}$, the densities of all species have the same expression in the \elasep{} as given 
in Table~\ref{table:cdmlasep}. In phase $\phase{0}$, the densities of $1$ and $\overline{1}$ are $(f(\lambda_{1}) - f(\lambda_{2}))$ and $(\overline{f}(\lambda_{1}) - f(\lambda_{2}))$ respectively.
We illustrate the density profiles with simulations for the $4$-species \elasep{} in Figure~\ref{fig:Bsimd}.

The shock picture in the \elasep{} is identical to that in the \olasep{} in all coexistence lines except the $\phase{0}-\phase{1}$
boundary.

On this coexistence line, $1$'s and $\overline{1}$'s form a shock with zero drift as shown in Figure~\ref{fig:gshockelasep}. The shock is pinned to the right (resp. left) boundary in phase $\phase{0}$ (resp. $\phase{1}$).  
In phase $\bphase{1}$, the density of these two species become equal and the height of the shock goes to zero as the system approaches this phase along the $\phase{0}-\phase{1}$ coexistence line.
We have performed simulations showing instantaneous density profiles for the \elasep{} with 4 species on the $\phase{0}-\phase{1}$ boundary and the results exactly match with the theoretical prediction.

\begin{figure}[htbp!]

	\begin{center} 
		\begin{tikzpicture}[scale=0.75]
		\draw[black,thick] (0,0) rectangle (10,10);
		\draw[black] (0,4)--(6,4)--(6,6)--(10,6);
		
		\draw[black] (0,1)--(10,1);\draw[black] (0,2)--(10,2);\draw[black] (0,7)--(10,7);
		\draw[black] (0,8)--(10,8);\draw[black] (0,9)--(10,9);\draw[black] (0,3)--(10,3);
		
		\node[] at (-1,4){$\scriptstyle f(\lambda_{1})$};\node[] at (-1,3){$\scriptstyle f(\lambda_{2})$};\node[] at (-1,1){$\scriptstyle f(\lambda_{r})$};
		\node[] at (11,6){$\scriptstyle \overline{f}(\lambda_{1})$};\node[] at (11,9){$\scriptstyle \overline{f}(\lambda_{r})$};\node[] at (11,7){$\scriptstyle \overline{f}(\lambda_{2})$};
		\node[] at (-1,2){$\vdots$};\node[] at (11,8){$\vdots$};
		
		\node[] at (5,7.5){$\overline{2}$}; \node[] at (5,9.5){$\overline{r}$}; \node[] at (5,0.5){$r$};\node[] at (5,2.5){$2$};
		\node[] at (5,8.6){$\vdots$};\node[] at (5,1.6){$\vdots$};
		\node[] at (3,5){$\overline{1}$};\node[] at (7.5,5){$1$};
		\draw[<->] (5.5,5)--(6.5,5);
		
		\node at (-1.8,5) {$\rho$};
		\node[] at (0,-0.5){$0$};\node[] at (10,-0.5){$1$};
		\node[] at (-0.5,0){$0$}; \node[] at (10.55,10){1};
		\node[] at (5,-1){$x$};

		\end{tikzpicture}		
	\end{center}
	\caption{The shock picture for the \elasep{} on the $\phase{0}-\phase{1}$ co-existence line.}
	\label{fig:gshockelasep}
\end{figure}


\begin{thebibliography}{10}

\bibitem{derrida1993}
B~Derrida, M~R Evans, V~Hakim, and V~Pasquier.
\newblock Exact solution of a 1d asymmetric exclusion model using a matrix
  formulation.
\newblock {\em Journal of Physics A: Mathematical and General}, 26(7):1493,
  1993.

\bibitem{blythe2007}
R~A Blythe and M~R Evans.
\newblock Nonequilibrium steady states of matrix-product form: a solver's
  guide.
\newblock {\em Journal of Physics A: Mathematical and Theoretical},
  40(46):R333, 2007.

\bibitem{schadschneider2002}
Andreas Schadschneider.
\newblock Traffic flow: a statistical physics point of view.
\newblock {\em Physica A: Statistical Mechanics and its Applications},
  313(1):153 -- 187, 2002.
\newblock Fundamental Problems in Statistical Physics.

\bibitem{chowdhury2000}
Debashish Chowdhury, Ludger Santen, and Andreas Schadschneider.
\newblock Statistical physics of vehicular traffic and some related systems.
\newblock {\em Physics Reports}, 329(4):199 -- 329, 2000.

\bibitem{penington2011}
Catherine~J. Penington, Barry~D. Hughes, and Kerry~A. Landman.
\newblock Building macroscale models from microscale probabilistic models: A
  general probabilistic approach for nonlinear diffusion and multispecies
  phenomena.
\newblock {\em Phys. Rev. E}, 84:041120, Oct 2011.

\bibitem{Evans1995}
M.~R. Evans, D.~P. Foster, C.~Godr{\`e}che, and D.~Mukamel.
\newblock Asymmetric exclusion model with two species: Spontaneous symmetry
  breaking.
\newblock {\em Journal of Statistical Physics}, 80(1):69--102, Jul 1995.

\bibitem{Arita2006b}
Chikashi Arita.
\newblock Phase transitions in the two-species totally asymmetric exclusion
  process with open boundaries.
\newblock {\em Journal of Statistical Mechanics: Theory and Experiment},
  2006(12):P12008, 2006.

\bibitem{uchiyama2008}
Masaru Uchiyama.
\newblock Two-species asymmetric simple exclusion process with open boundaries.
\newblock {\em Chaos, Solitons and Fractals}, 35(2):398 -- 407, 2008.

\bibitem{ayyer2009}
Arvind Ayyer, Joel~L. Lebowitz, and Eugene~R. Speer.
\newblock On the two species asymmetric exclusion process with semi-permeable
  boundaries.
\newblock {\em Journal of Statistical Physics}, 135(5):1009--1037, Jun 2009.

\bibitem{ayyer2012}
Arvind Ayyer, Joel~L. Lebowitz, and Eugene~R. Speer.
\newblock On some classes of open two-species exclusion processes.
\newblock {\em Markov Processes And Related Fields}, 18(5):157--176, 2012.

\bibitem{crampe2015}
N~Crampe, K~Mallick, E~Ragoucy, and M~Vanicat.
\newblock Open two-species exclusion processes with integrable boundaries.
\newblock {\em Journal of Physics A: Mathematical and Theoretical},
  48(17):175002, 2015.

\bibitem{crampe-evans-mallick-ragoucy-vanicat2016}
N~Crampe, M~R Evans, K~Mallick, E~Ragoucy, and M~Vanicat.
\newblock Matrix product solution to a 2-species {TASEP} with open integrable
  boundaries.
\newblock {\em Journal of Physics A: Mathematical and Theoretical},
  49(47):475001, 2016.

\bibitem{ayyer2018}
Arvind Ayyer, Caley Finn, and Dipankar Roy.
\newblock Matrix product solution of a left-permeable two-species asymmetric
  exclusion process.
\newblock {\em Phys. Rev. E}, 97:012151, Jan 2018.

\bibitem{evans-ferrari-mallick-2009}
Martin~R. Evans, Pablo~A. Ferrari, and Kirone Mallick.
\newblock Matrix representation of the stationary measure for the multispecies
  {TASEP}.
\newblock {\em Journal of Statistical Physics}, 135(2):217--239, Apr 2009.

\bibitem{prolhac-evans-mallick-2009}
S~Prolhac, M~R Evans, and K~Mallick.
\newblock The matrix product solution of the multispecies partially asymmetric
  exclusion process.
\newblock {\em Journal of Physics A: Mathematical and Theoretical},
  42(16):165004, 2009.

\bibitem{FM06}
Pablo~A. Ferrari and J.~B. Martin.
\newblock Multi-class processes, dual points and {M}/{M}/1 queues.
\newblock {\em Markov Process. Related Fields}, 12(2):175--201, 2006.

\bibitem{FM07}
Pablo~A. Ferrari and James~B. Martin.
\newblock Stationary distributions of multi-type totally asymmetric exclusion
  processes.
\newblock {\em The Annals of Probability}, 35(3):807--832, 05 2007.

\bibitem{ayyer-linusson-2016}
Arvind Ayyer and Svante Linusson.
\newblock Correlations in the multispecies {TASEP} and a conjecture by lam.
\newblock {\em Trans. Amer. Math. Soc.}, 369(2):1097--1125, 2017.

\bibitem{cantini2016}
Luigi Cantini, Alexandr Garbali, Jan de~Gier, and Michael Wheeler.
\newblock Koornwinder polynomials and the stationary multi-species asymmetric
  exclusion process with open boundaries.
\newblock {\em Journal of Physics A: Mathematical and Theoretical},
  49(44):444002, 2016.

\bibitem{ayyer2017}
Arvind Ayyer and Dipankar Roy.
\newblock The exact phase diagram for a class of open multispecies asymmetric
  exclusion processes.
\newblock {\em Scientific Reports}, 7:13555--, Oct 2017.

\bibitem{crampe2016}
N~Crampe, C~Finn, E~Ragoucy, and M~Vanicat.
\newblock Integrable boundary conditions for multi-species {ASEP}.
\newblock {\em Journal of Physics A: Mathematical and Theoretical},
  49(37):375201, 2016.

\bibitem{uchiyama2004}
Masaru Uchiyama, Tomohiro Sasamoto, and Miki Wadati.
\newblock Asymmetric simple exclusion process with open boundaries and
  {A}skey--{W}ilson polynomials.
\newblock {\em Journal of Physics A: Mathematical and General}, 37(18):4985,
  2004.

\end{thebibliography}
\end{document}